\documentclass[aps,prb,amsmath,amssymb,reprint,twocolumn]{revtex4-2}
\usepackage{graphicx}
\usepackage{epstopdf}
\usepackage{amsmath}
\usepackage{indentfirst}
\usepackage{bm}
\usepackage[utf8]{inputenc}
\usepackage[T1]{fontenc}

\begin{document}
	\title{Exceptional Points in a Parallel Double-Quantum-Dot Josephson Junction Coupled to a Ferromagnetic Reservoir}
	
	\author{Yiyan Wang}
	\affiliation{Key Laboratory of Artificial Structures and Quantum Control (Ministry of Education), Department of Physics and 
		Astronomy, Shanghai Jiaotong University, 800 Dongchuan Road, Shanghai 200240, China}
	\author{Ruixin Zhou}
	\affiliation{Key Laboratory of Artificial Structures and Quantum Control (Ministry of Education), Department of Physics and 
		Astronomy, Shanghai Jiaotong University, 800 Dongchuan Road, Shanghai 200240, China}
		
	\author{Bing Dong}
	\thanks{Author to whom correspondence should be addressed. Email:bdong@sjtu.edu.cn.}
	\affiliation{Key Laboratory of Artificial Structures and Quantum Control (Ministry of Education), Department of Physics and 
		Astronomy, Shanghai Jiaotong University, 800 Dongchuan Road, Shanghai 200240, China}
	
	\begin{abstract}
		We investigate exceptional points (EPs) in a parallel double-quantum-dot Josephson junction coupled to a dissipative reservoir. By integrating out the leads, we obtain a non-Hermitian Bogoliubov-de Gennes description wherein the superconducting phase difference and orbital flux govern the complex Andreev spectrum. For spin-independent dissipation, the second-order EPs identified within the infinite superconducting gap limit are eliminated when the finite superconducting gap is properly incorporated. In contrast, spin-dependent dissipation originating from a ferromagnetic reservoir, in conjunction with magnetic flux, gives rise to second-order EPs that persist in superconducting leads with finite gap. Moreover, flux tuning enables the coalescence of two second-order EPs into a third-order EP, whose eigenvalue splitting exhibits cubic-root scaling behavior. A many-body parity analysis establishes the connection between the contrasting finite-gap behavior and the spectral relationship between the even- and odd-parity sectors. Finally, Josephson currents calculated from both the free-energy derivative and the surrogate-model density matrix demonstrate consistency and remain continuous across the EPs. These findings establish spin-selective dissipation and interferometric flux as effective control parameters for robust non-Hermitian singularities in superconducting nanostructures.
	\end{abstract}
	\maketitle
	\section{Introduction}
	Non-Hermitian physics provides a framework for describing open systems with dissipation, gain, and reservoir-induced finite lifetimes~\cite{Rotter2009,Ashida2021,Bergholtz2021,Okuma2023,Ding2022,KoziiFu2024}. A central feature is the exceptional point (EP), where both eigenvalues and eigenvectors coalesce and the Hamiltonian becomes defective~\cite{Heiss2012,Kawabata2019,Ding2022}. Unlike ordinary Hermitian degeneracies, EPs are branch-point singularities of complex spectra and may carry topological charges associated with point-gap winding~\cite{Kawabata2019,Bessho2020,Bergholtz2021}. Their topology and nonadiabatic encircling have been observed in microwave and optomechanical experiments and explored extensively in photonic systems~\cite{Dembowski2001,Xu2016,Doppler2016,ElGanainy2018,Feng2017,Miri2019}.
	
	Superconducting mesoscopic devices provide a natural platform for exploring such non-Hermitian spectral singularities. 
	In Josephson junctions, the low-energy physics is governed by Andreev bound states (ABSs)~\cite{Beenakker1991,MartinRodero2011}, whose energies are controlled by superconducting phase differences, magnetic fluxes, gate voltages, and tunnel couplings~\cite{Pillet2010,Hays2018}.
	Quantum-dot Josephson junctions are particularly valuable since their spectrum and supercurrent can be engineered through local gates and tunnel barriers~\cite{vanDam2006,Eichler2007,Karrasch2008,Zitko2010,Choi2000}. 
	Ferromagnetic coupling further introduces spin-dependent proximity and relaxation channels~\cite{Buzdin2005,Bergeret2005,Linder2015}. 
	Coupling the junction to normal-metal or ferromagnetic reservoirs endows these states with finite lifetimes and generates effective non-Hermitian self-energies~\cite{Cayao2024PhaseBiased,Capecelatro2025}; The resulting complex ABS spectrum can therefore host singular degeneracies of the non-Hermitian Bogoliubov-de Gennes (BdG) Hamiltonian~\cite{Cayao2024PhaseBiased,Ohnmacht2025}.
	
	Non-Hermitian Green-function methods have recently been applied to dissipative Josephson transport, with EPs or exceptional rings predicted in phase-biased and multiterminal junctions~\cite{Shen2024,Capecelatro2025,Pino2025,Cayao2024MultiTerminal,Ohnmacht2025,Solow2025}.  
	An Andreev-only Hamiltonian is reliable when the relevant poles remain well-separated from the near-gap continuum~\cite{Capecelatro2025}.  
	However, many quantum-dot treatments additionally assume an infinite superconducting gap and hence a static induced pairing. 
	Whether EPs found in that limit persist after finite-gap quasiparticle degrees of freedom are restored is therefore an open question.
	
	Parallel double-dot junctions enable independent manipulation of local and nonlocal pairing through superconducting phase and orbital-flux interference~\cite{Probst2016,Tomaszewski2018,Choi2000,Zitko2010}. 
	In this work, we investigate such a junction coupled to either a normal or ferromagnetic reservoir and compare its infinite-gap BdG spectrum with a finite-level superconducting surrogate calculation~\cite{Li2025QPT,Wang2025Flux}. 
	Spin-independent dissipation generates second-order EPs (EP2s) in the static model, which are eliminated in the finite-gap calculation for the considered parameters. 
	In contrast, spin-selective dissipation combined with magnetic flux preserves a subset of EP2s in the surrogate model. 
	Furthermore, flux tuning facilitates the coalescence of two EP2s into a third-order EPs (EP3) within the infinite-gap model.
	However, finite-gap corrections modify the degeneracy conditions, preventing the same parameter values from achieving this coalescence.
	
	We further compare Matsubara and biorthogonal-density-matrix calculations of the Josephson current and find consistent, continuous current--phase relations across the surviving EP2s. A parity-resolved Fock-space analysis is used as a diagnostic of the contrasting finite-gap behavior.
	
	\section{Model Hamiltonian and Theoretical Method}
	In this work, we investigate a hybrid nanodevice consisting of two quantum dots (QDs) connected in parallel between two superconducting (SC) leads, with a magnetic flux threading through the loop. One of the QDs is coupled to either a normal or ferromagnetic lead, as depicted in Fig. 1. The system Hamiltonian is given by:
	\begin{equation}
		H=H_{0}+H_{\mathrm{I}},
	\end{equation}
	with
	\begin{equation}
		\begin{split}
			H_{0}=&\sum_{j}\varepsilon_{j\sigma}d^{\dag}_{j\sigma}d_{j\sigma}
			+\sum_{\sigma}\left(t_{d}d^{\dag}_{1\sigma}d_{2\sigma}+h.c.\right)\\
			+&\sum_{\eta \bm{k} \sigma}\varepsilon_{\eta \bm{k} \sigma}c_{\eta \bm{k} \sigma}^{\dag}c_{\eta \bm{k} \sigma}+\sum_{\eta \bm{k}}\left(\Delta e^{\mathrm{i}\phi_{\eta}}c_{\eta \bm{k} \uparrow}^{\dag}c_{\eta -\bm{k} \downarrow}^{\dag}+h.c.\right)\\
			+&\sum_{\bm{k'}\sigma}\varepsilon_{\bm{k'}\sigma} c^{\dag}_{\bm{k'}\sigma} c_{\bm{k'}\sigma},
		\end{split}
	\end{equation}
	\begin{equation}
		H_{\mathrm{I}}=\sum_{j \eta \bm{k} \sigma}(V_{\eta j} c_{\eta \bm{k} \sigma}^{\dag}d_{j\sigma}+h.c.)+\sum_{\bm{k'} \sigma}(W_{\sigma} c_{\bm{k'} \sigma}^{\dag}d_{1\sigma}+h.c.),
	\end{equation}	
	Here, $H_0$ represents the Hamiltonian in the absence of tunneling, where $d^{\dag}_{j\sigma}$ creates an electron with spin $\sigma$ and energy $\varepsilon_{j \sigma}$ on the $j$-th ($j=1,2$) quantum dot. 
	For simplicity, we neglect Zeeman splitting and assume $\varepsilon_{j\uparrow}=\varepsilon_{j\downarrow}=\varepsilon_j$.
	The parameter $t_d$ denotes the inter-dot hopping amplitude. 
	The operator $c_{\eta \bm{k} \sigma}^{\dag}$ creates an electron with spin $\sigma$, momentum $\bm{k}$, and energy $\varepsilon_{\eta \bm{k} \sigma}$ in the $\eta$-th superconducting lead ($\eta=\mathrm{L},\mathrm{R}$).
	In this context, $\phi_\eta$ and $\Delta$ represent the superconducting phase and pair potential, respectively.
	Additionally, $c_{\bm{k'}\sigma}^{\dag}$ creates an electron with spin $\sigma$, momentum $\bm{k'}$, and energy $\varepsilon_{\bm{k'} \sigma}$ in the normal or ferromagnetic lead. 
	The term $H_\mathrm{I}$ describes the tunneling between the double quantum dot system and the two superconducting leads along with the normal/ferromagnetic lead, characterized by the corresponding tunneling amplitudes $V_{\eta j}$ and $W_\sigma$. The tunneling amplitude $V_{\eta j}$ is expressed as:
	\begin{equation}
		\begin{split}
			V_{\mathrm{L1}}=V \mathrm{e}^{-\mathrm{i}\frac{\phi_{\mathbf{B}}}{4}},\, 
			V_{\mathrm{R1}}=V \mathrm{e}^{\mathrm{i}\frac{\phi_{\mathbf{B}}}{4}},\\
			V_{\mathrm{L2}}=V \mathrm{e}^{\mathrm{i}\frac{\phi_{\mathbf{B}}}{4}},\, 
			V_{\mathrm{R2}}=V \mathrm{e}^{-\mathrm{i}\frac{\phi_{\mathbf{B}}}{4}},
		\end{split}
	\end{equation}
	where $\phi_\mathbf{B}=2\pi\Phi_{\mathbf{B}_0}/\Phi_0$ represents the Aharonov-Bohm phase associated with the magnetic flux threading the closed loop.
	Here, $\Phi_0=hc/e$ denotes the single-electron flux quantum (equivalent to twice the superconducting flux quantum $hc/2e$).  
	This Peierls-phase formulation accurately describes the Aharonov-Bohm interference arising from the two parallel quantum dot paths~\cite{Tomaszewski2018}. 
	The parameters $W_{\sigma}$ characterize the tunneling amplitudes between QD1 and the normal or ferromagnetic lead, with $W_{\uparrow}=W_{\downarrow}=W$ for the normal lead case, while $W_{\uparrow}\neq W_{\downarrow}$ for the ferromagnetic lead configuration. 
	Throughout this work, we employ natural units where $\mathrm{\hbar}= k_\mathrm{B} = e = 1$.
	
	We formulate the theoretical framework using a Grassmann path integral approach and integrate out the noninteracting leads, following the standard Green's function construction for open Josephson junctions~\cite{Capecelatro2025, Shen2024}. 
	Within the wide-band limit, the superconducting hybridization strength and spin-dependent dissipation rates are given by $\Gamma=2\pi\rho_{\mathrm{SC}}V^2$ and $\gamma_{\uparrow/\downarrow}=\pi\rho_{\mathrm{N}}W^2_{\uparrow/\downarrow}$, respectively. 
	In these expressions, $\rho_{\mathrm{SC}}$ and $\rho_{\mathrm{N}}$ represent the normal-state densities of states of the superconducting and normal/ferromagnetic reservoirs.
	
	\begin{figure}[tbp]
		\centering
		\includegraphics[width=0.9\linewidth]{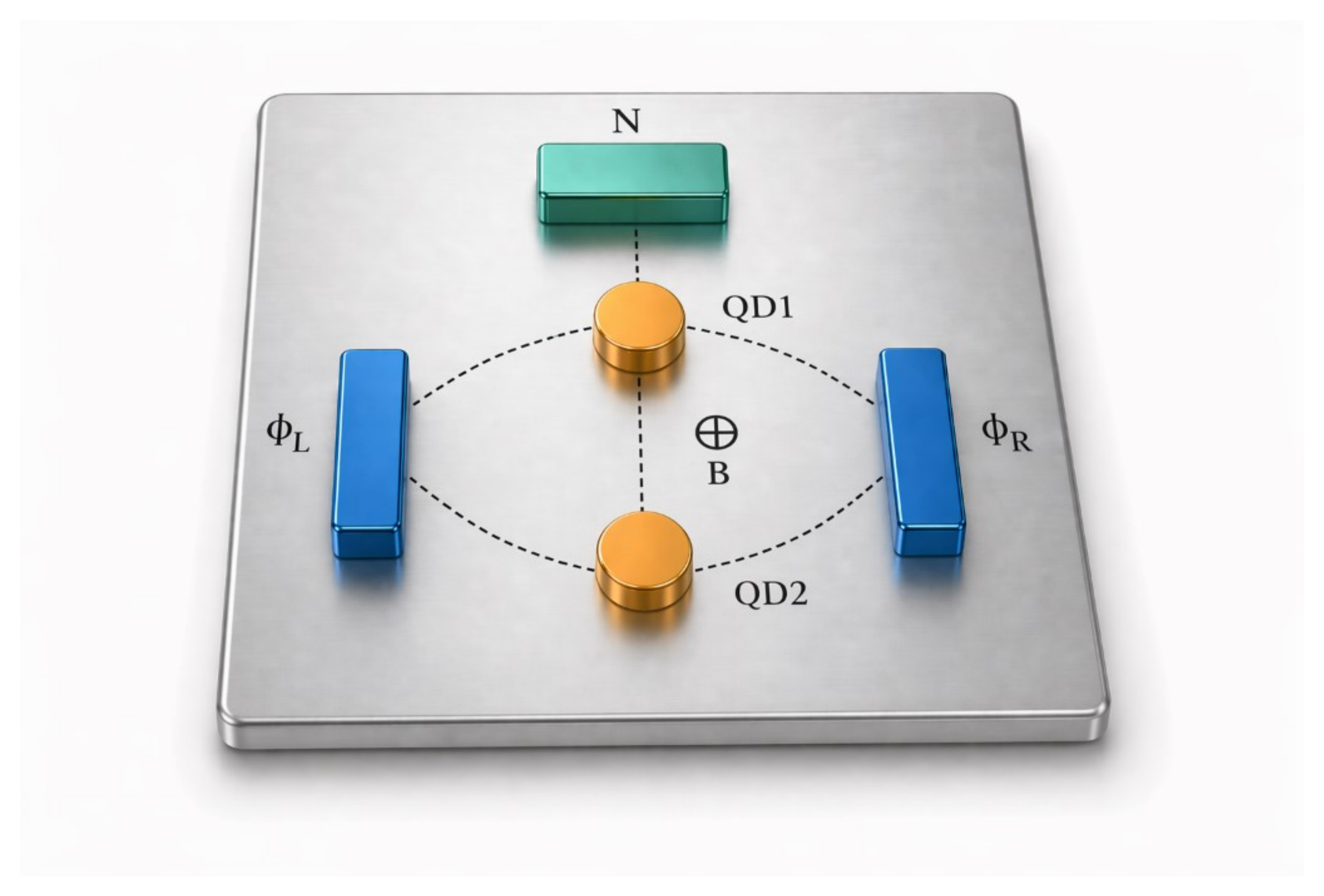}
		\caption{(Color online) Parallel double-quantum-dot Josephson junction coupled to a normal or ferromagnetic reservoir.}
		\label{fig_device}
	\end{figure}
	
	Performing Fourier transformation in imaginary time yields the fermionic Matsubara frequencies $\omega_n=(2n+1)\pi/\beta$, $n\in\mathbb{Z}$, where $\beta=1/(k_\mathrm{B}T)$. The effective action $S_{\text{eff}}$ takes the form:
	\begin{equation}
		S_\text{eff}=\sum_{\omega_{n}}\bar{\psi_d} \mathcal{M}\left(\omega_{n}\right) \psi_d,
	\end{equation}
	with the matrix $\mathcal{M}\left(\omega_n\right)$ given by:
	\begin{align}
		\mathcal{M}\left(\omega_n\right)=&-\mathrm{i}\omega_{n}+\frac{1}{2}\left(\varepsilon_{1}+\varepsilon_2\right)\sigma_z\tau_0+\frac{1}{2}\left(\varepsilon_{1}-\varepsilon_2\right)\sigma_z\tau_z\\
		&+t_{d}\sigma_z\tau_x-\Sigma_{\text{SC}}\left(\omega_{n}\right)-\Sigma_{\text{N}}\left(\omega_{n}\right),
	\end{align}
	where $\sigma_i$ and $\tau_i$ denote the Pauli matrices in Nambu and site representations, respectively. We define $\delta\phi=\phi_{\mathrm{L}}-\phi_{\mathrm{R}}$ as the phase difference between the left and right superconducting leads. The self-energies $\Sigma_{\text{SC}}\left(\omega_{n}\right)$ and $\Sigma_{\text{N}}\left(\omega_{n}\right)$ for the superconducting and normal/ferromagnetic leads, respectively, are given by:
	\begin{widetext}
		\begin{equation}
				\Sigma_{\text{SC}}\left(\omega_n\right)=\frac{2}{\pi}\frac{\arctan\left(\frac{D}{\sqrt{\Delta^2+\omega_{n}^2}}\right)}{\sqrt{\Delta^2+\omega_{n}^2}}\times\Gamma\renewcommand{\arraystretch}{0.1}
				\begin{bmatrix}
					\mathrm{i}\omega_{n}&\Delta \cos\left(\frac{\delta\phi-\phi_{\mathbf{B}}}{2}\right)&\mathrm{i}\omega_{n}\mathrm{e}^{\mathrm{i}\frac{\phi_{\mathbf{B}}}{2}}&\Delta \cos\left(\frac{\delta\phi}{2}\right)\\
					\Delta \cos\left(\frac{\delta\phi-\phi_{\mathbf{B}}}{2}\right)&\mathrm{i}\omega_{n}&\Delta \cos\left(\frac{\delta\phi}{2}\right)&\mathrm{i}\omega_{n}\mathrm{e}^{-\mathrm{i}\frac{\phi_{\mathbf{B}}}{2}}\\
					\mathrm{i}\omega_{n}\mathrm{e}^{-\mathrm{i}\frac{\phi_{\mathbf{B}}}{2}}&\Delta \cos\left(\frac{\delta\phi}{2}\right)&\mathrm{i}\omega_{n}&\Delta \cos\left(\frac{\delta\phi+\phi_{\mathbf{B}}}{2}\right)\\
					\Delta \cos\left(\frac{\delta\phi}{2}\right)&\mathrm{i}\omega_{n}\mathrm{e}^{\mathrm{i}\frac{\phi_{\mathbf{B}}}{2}}&\Delta \cos\left(\frac{\delta\phi+\phi_{\mathbf{B}}}{2}\right)&\mathrm{i}\omega_{n}
				\end{bmatrix},
		\end{equation}
	\end{widetext}
	\begin{equation}
		\Sigma_{\text{N}}\left(\omega_n\right)=\begin{bmatrix}
			-\mathrm{i}\gamma_\uparrow\,\mathrm{sgn}\,\omega_n & 0 & 0 & 0 \\
			0 & -\mathrm{i}\gamma_\downarrow\,\mathrm{sgn}\,\omega_n & 0 & 0 \\
			0 & 0 & 0 & 0 \\
			0 & 0 & 0 & 0 
		\end{bmatrix}.
	\end{equation}
	After integrating out the Grassmann fields $\left\{\psi_d,\bar{\psi}_d\right\}$, the regularized functional determinant is expressed as:
	\begin{equation}
		\ln Z=\sum_{\omega_n}
		\left[\ln\det\mathcal{M}(\omega_n)-\ln\det\mathcal{M}_0(\omega_n)\right],
		\label{Matsubara_lnZ}
	\end{equation}
	where $\mathcal{M}_0$ is obtained by setting $\Delta=0$ in the superconducting self-energy and serves as the normal-state reference. 
	The free energy is given by:
	\begin{equation}
		F=-\frac{1}{\beta}\ln Z.
		\label{Matsubara_free_energy}
	\end{equation}
	To obtain the retarded spectrum, we perform analytic continuation of the Matsubara kernel via $\mathrm{i}\omega_n\rightarrow\omega+\mathrm{i}0^+$. Subsequently, we take the limit $\Delta\to\infty$.
	In this regime, the superconducting self-energy becomes frequency-independent, and the inverse retarded Green's function defines the following low-energy effective Hamiltonian:
	\begin{equation}
		H_{\text{eff}}=\psi_d^\dagger \mathcal{H}\psi_d,
	\end{equation}
	where
	\begin{widetext}
		\begin{equation}
			\mathcal{H}=
			\begin{bmatrix}
				\varepsilon_1-\mathrm{i}\gamma_\uparrow&\Gamma\cos\left(\frac{\delta\phi-\phi_\mathbf{B}}{2}\right)&t_d&\Gamma\cos\left(\frac{\delta\phi}{2}\right)\\
				\Gamma\cos\left(\frac{\delta\phi-\phi_\mathbf{B}}{2}\right)&-\varepsilon_1-\mathrm{i}\gamma_\downarrow&\Gamma\cos\left(\frac{\delta\phi}{2}\right)&-t_d\\
				 t_d&\Gamma\cos\left(\frac{\delta\phi}{2}\right)&\varepsilon_2&\Gamma\cos\left(\frac{\delta\phi+\phi_\mathbf{B}}{2}\right)\\
				\Gamma\cos\left(\frac{\delta\phi}{2}\right)&-t_d&\Gamma\cos\left(\frac{\delta\phi+\phi_\mathbf{B}}{2}\right)&-\varepsilon_2\\
			\end{bmatrix}.\label{Non-Hermitian Low Energy Effective H}
		\end{equation}
	\end{widetext}
	We set $\varepsilon_j=0$ to focus on the EPs. 
	Since the wide-band normal-lead self-energy depends solely on $\operatorname{sgn}\omega_n$, we employ the fitted finite representation of the superconducting-lead dynamics obtained from the imaginary-time formulation, while maintaining the explicit treatment of the dissipative dot self-energy. 
	The resulting surrogate Hamiltonian is expressed as:
	\begin{equation}
		\tilde{H}=\tilde{H}_{\mathrm{DQD}}+\tilde{H}_{\mathrm{SC}}+\tilde{H}_\mathrm{T},
		\label{Non-Hermitian Surrogate_H}
	\end{equation}
	where
	\begin{equation}
		\begin{split}
			\tilde{H}_{\mathrm{DQD}}=&-\mathrm{i}\gamma_\uparrow d^{\dag}_{1\uparrow}d_{1\uparrow}-\mathrm{i}\gamma_\downarrow d^{\dag}_{1\downarrow}d_{1\downarrow}\\
			&+\sum_{j\sigma}\varepsilon_j d_{j\sigma}^\dagger d_{j\sigma} 
			+\sum_{\sigma}\left(t_{d}d^{\dag}_{1\sigma} d_{2\sigma}+h.c.\right),
		\end{split}
	\end{equation}
	\begin{equation}
		\tilde{H}_{\mathrm{SC}}=\sum_{\eta l \sigma}\tilde{\xi}_{l}c_{\eta l \sigma}^{\dag}c_{\eta l \sigma}+\sum_{\eta l}(\Delta e^{\mathrm{i}\phi_{\eta}}c_{\eta l \uparrow}^{\dag}c_{\eta l \downarrow}^{\dag}+h.c.),
	\end{equation}
	\begin{equation}
		\tilde{H}_{\mathrm{T}}=\sum_{i\eta  l \sigma}(\tilde{V}_{\eta i l} c_{\eta l \sigma}^{\dag}d_{i\sigma}+h.c.).
	\end{equation}
	For the spectral calculation, we utilize the following five-point parameter set for each superconducting lead:
	\begin{equation}
		\begin{split}
			\tilde\xi_l/\Delta={}&\{9.233388,1.64236,0,\\
				&-1.64236,-9.233388\},\\
			\gamma_l/\Delta={}&\{6.53577,0.840433,0.3875,\\
				&0.840433,6.53577\}.
		\end{split}
	\end{equation}
	The parameters $\tilde\xi_l$ and $\gamma_l$ are derived from the shape-factor fit employed in the imaginary-time free-energy calculation ~\cite{Li2025QPT, Wang2025Flux}, with bandwidth $D=100,\Delta$; these parameters are not independently adjusted in the spectral calculation.
	The tunneling amplitudes are given by $\tilde t_l=\sqrt{\gamma_l\Gamma/2}$.
	In the absence of Coulomb interactions, the resulting surrogate Hamiltonian can be formulated in BdG form and diagonalized directly.
	
	\section{Results and Discussions}
	
	\subsection{Spectrum}
	We first examine the spin-independent case, $\gamma_\uparrow=\gamma_\downarrow=\gamma$, at $\phi_{\mathbf{B}}=0$, $t_d=0$, and $\varepsilon_1=\varepsilon_2=0$. The eigenvalues of Eq.~\eqref{Non-Hermitian Low Energy Effective H} are
	\begin{equation}
		E=-\frac{\mathrm{i}\gamma}{2}
		\pm\Gamma\cos\left(\frac{\delta\phi}{2}\right)
		\pm\frac{\mathrm{i}}{2}\sqrt{\gamma^2-4\Gamma^2\cos^2\left(\frac{\delta\phi}{2}\right)}.\label{RootEQ_1}
	\end{equation}
	Figure~2(a) shows the corresponding complex spectrum.
	
		\begin{figure*}[htbp]
			\begin{minipage}{0.75\linewidth}
				\centering
				\includegraphics[width=0.45\linewidth]{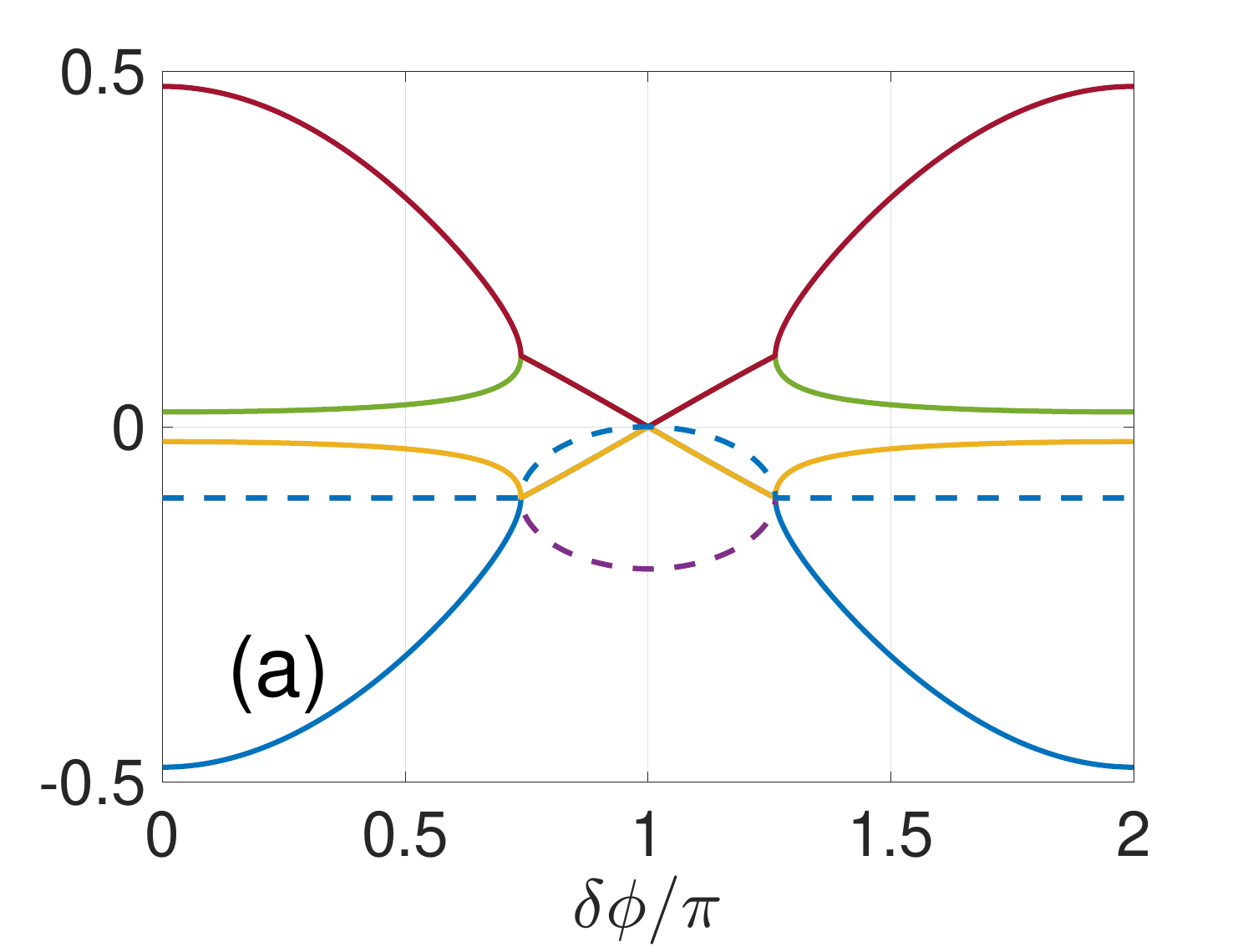}
				\includegraphics[width=0.45\linewidth]{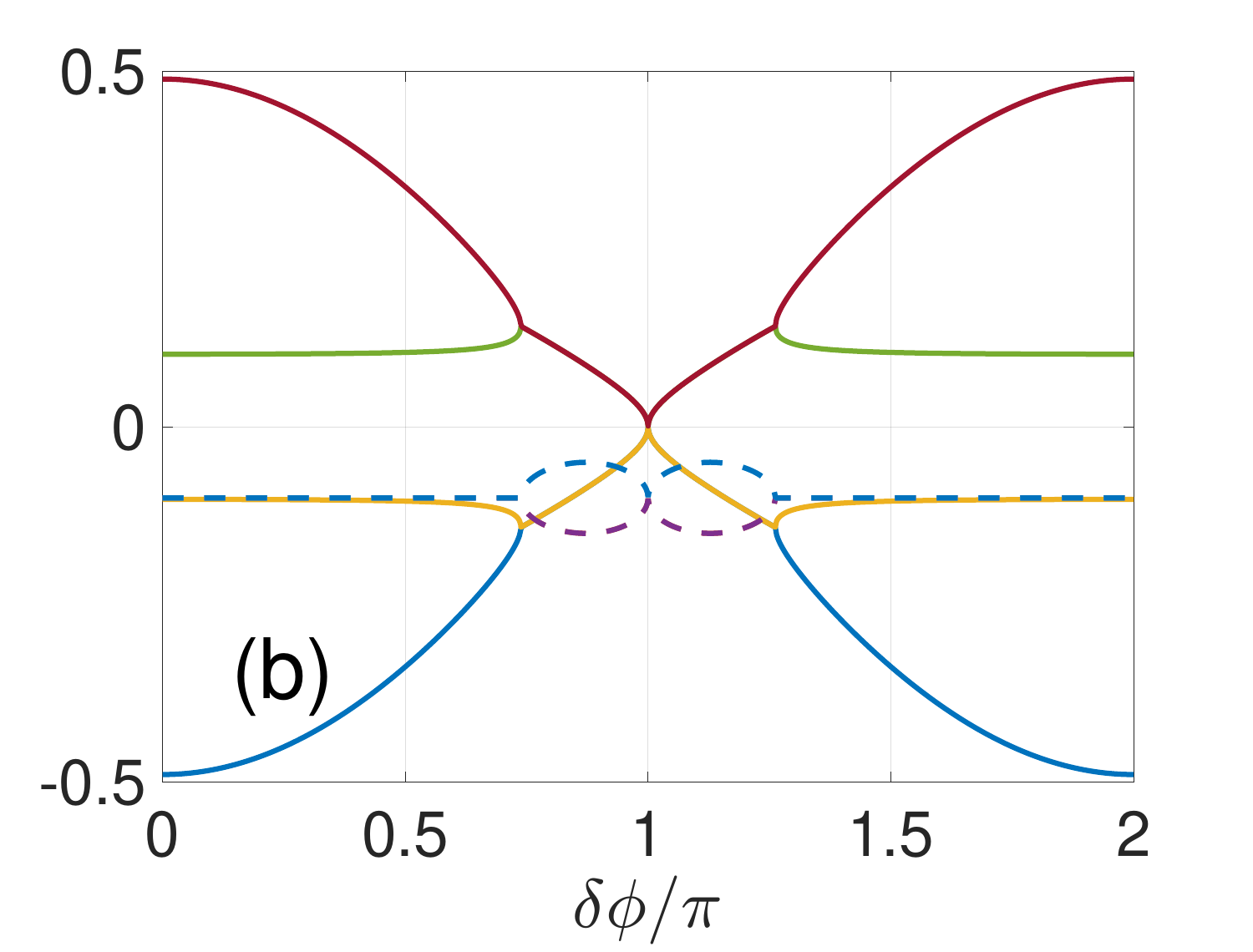}
			\end{minipage}
			\centering
			\begin{minipage}{0.75\linewidth}
				\includegraphics[width=0.45\linewidth]{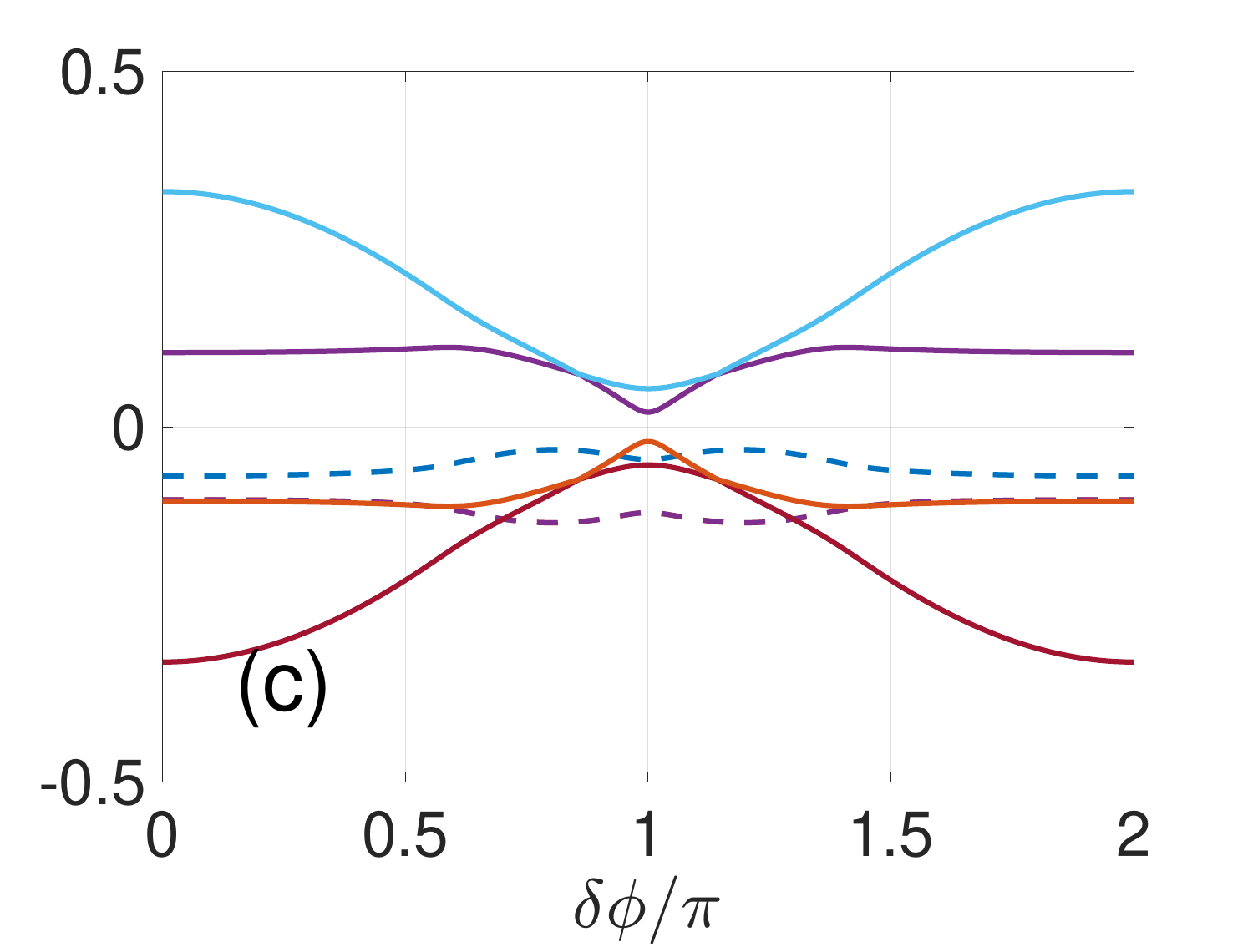}
				\includegraphics[width=0.45\linewidth]{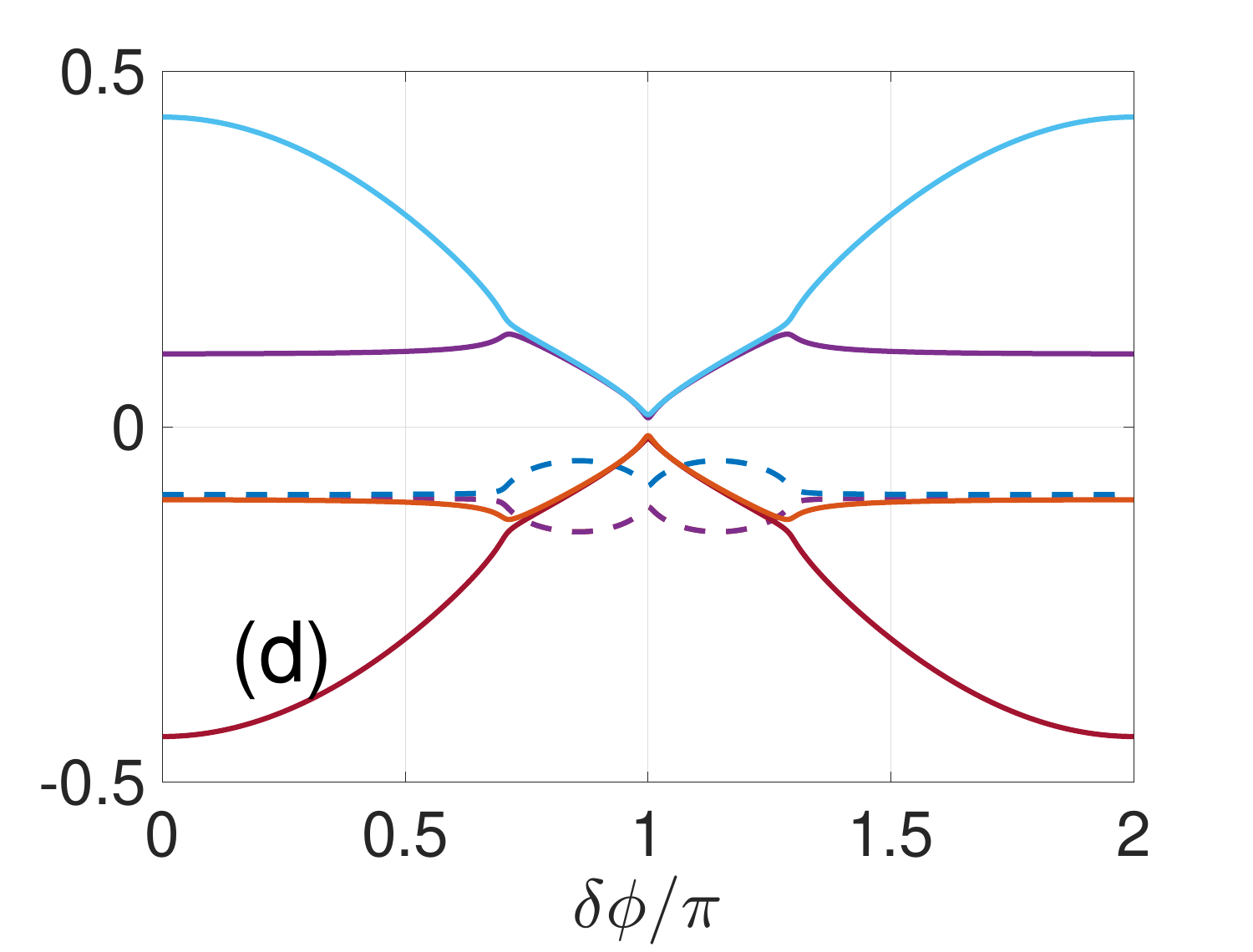}
			\end{minipage}
			\caption{(Color online) Energy spectrum of the parallel double quantum dot system based on the BdG Hamiltonian. The solid and dashed lines represent the real and imaginary parts of the eigenenergies, respectively. Parameters are set as: $\Gamma = 0.25$, $\gamma_\uparrow = \gamma_\downarrow = 0.2$, $\phi_\mathbf{B} = 0$ and (a) $\Delta\to\infty$, $t_d = 0$. (b) $\Delta\to\infty$, $t_d = 0.1$. (c) $\Delta=1$, $t_d = 0.1$. (d)$\Delta=10$, $t_d = 0.1$.}
		\end{figure*}

	At $\delta\phi=2\arccos[\gamma/(2\Gamma)]$, the square-root discriminant in Eq.~\eqref{RootEQ_1} vanishes, indicating the presence of EP2 characterized by the coalescence of both eigenvalues and eigenvectors.
	When the interdot hopping is set to $t_d=\gamma/2$, all four eigenvalues converge at $\delta\phi=\pi$, although the defectiveness remains of second order. 
	quartic roots are then given by:
	\begin{widetext}
		\begin{equation}
			E=\frac{\mathrm{i}}{2}\left(-\gamma \pm
			\sqrt{-4 t_d^2 + \gamma^2 - 
				4 \Gamma^2 \left(1 + \cos\delta\phi\right) \pm
				4 \mathrm{i} \sqrt{\Gamma^2 \cos\left(\frac{\delta\phi}{2}\right)^2\left(-2 \Gamma^2 + \gamma^2 - 
					2 \Gamma^2 \cos\delta\phi\right)}}\right).
		\end{equation}
	\end{widetext}
	This spectrum is displayed in Fig.~2(b). 
	Consequently, algebraic multiplicity alone does not determine the EP order; rather, the order is established by the coalescing eigenvectors (equivalently, by the corresponding Jordan block structure).
	
	We proceed to diagonalize the finite-gap surrogate Hamiltonian presented in Eq.~\eqref{Non-Hermitian Surrogate_H}. 
	For $\Delta=1$ [Fig.~2(c)], the eigenvalue coalescences characteristic of the infinite-gap model are replaced by avoided crossings in the surrogate spectrum.
	Increasing the gap parameter restores the large-gap behavior [Fig.~2(d)]. 
	Consequently, for these parameters, the five-level representation indicates that finite-$\Delta$ energy dependence eliminates the exact coalescences predicted by the static-pairing approximation.   
	A direct continuous-self-energy pole calculation would be necessary to establish the corresponding quantitative statement for the original reservoir system.
	
	We now introduce spin-selective dissipation with $\gamma_\uparrow=0.75$ and $\gamma_\downarrow=0.25$, corresponding to a ferromagnetic reservoir. 
	With $\phi_{\mathbf{B}}=0.5\pi$ and $t_d=0$, the magnetic flux and spin-dependent loss mechanisms jointly govern the spectrum displayed in Fig.~3(a).
	
	\begin{figure*}[htbp]
		\begin{minipage}{0.86\linewidth}
			\centering
			\includegraphics[width=0.46\linewidth]{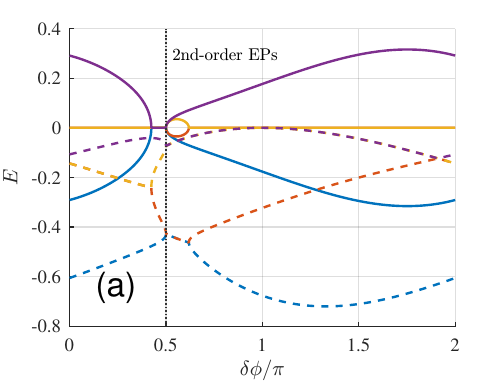}
			\hspace{0.035\linewidth}
			\includegraphics[width=0.46\linewidth]{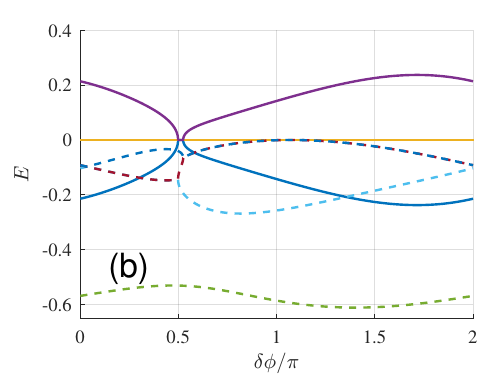}
		\end{minipage}
		\caption{(Color online) Energy spectrum of the parallel double quantum dot (DQD) system based on the BdG Hamiltonian. The solid and dashed lines represent the real and imaginary parts of the eigenenergies, respectively. Parameters are set as: $\Gamma = 0.25$, $\gamma_\uparrow = 0.75, \gamma_\downarrow = 0.25$, $\phi_\mathbf{B} = 0.5\,\pi$, $t_d=0$ and (a) $\Delta\to\infty$, (b) $\Delta=1$.}
	\end{figure*}
	
	The infinite-gap spectrum exhibits a pair of coincident second-order EPs at $\delta\phi=0.497772\pi$ and additional EPs at other phase biases. 
	Although four eigenvalue branches converge at the coincident point, each singularity maintains second-order character. 
	In the finite-gap calculation at $\Delta=1$, two EPs are lifted while two second-order EPs persist [Fig.~3(b)]. 
	Consequently, a subset of the EPs survives in the five-level finite-gap surrogate model with spin-selective dissipation.
	
	The orbital flux also serves as a continuous control parameter for the EP positions. We set $\phi_{\mathbf{B}}=0.24075202\pi$ and maintain the remaining parameters from Fig.~3(a); the resulting spectrum is displayed in Fig.~4(a).
	
	\begin{figure*}[htbp]
		\begin{minipage}{1.0\linewidth}
			\centering
			\begin{minipage}[c]{0.36\linewidth}
				\centering
				\includegraphics[width=\linewidth]{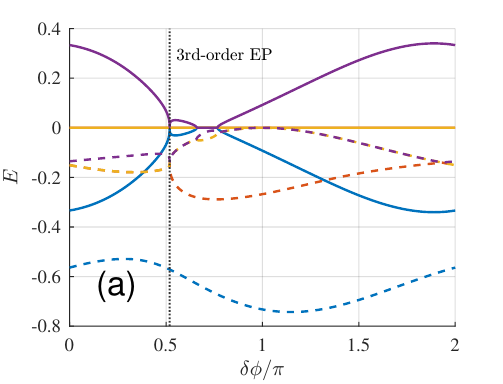}
			\end{minipage}\hspace{0.025\linewidth}%
			\begin{minipage}[c]{0.36\linewidth}
				\centering
				\includegraphics[width=\linewidth]{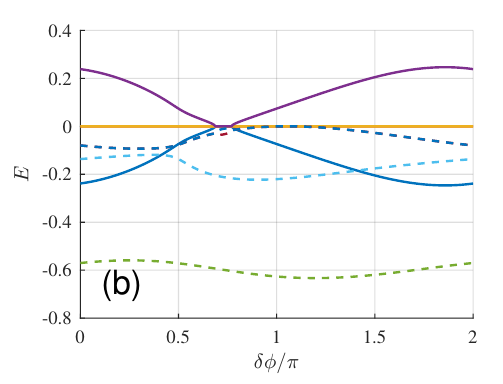}
			\end{minipage}\\[-0.5ex]
			\includegraphics[width=0.82\linewidth]{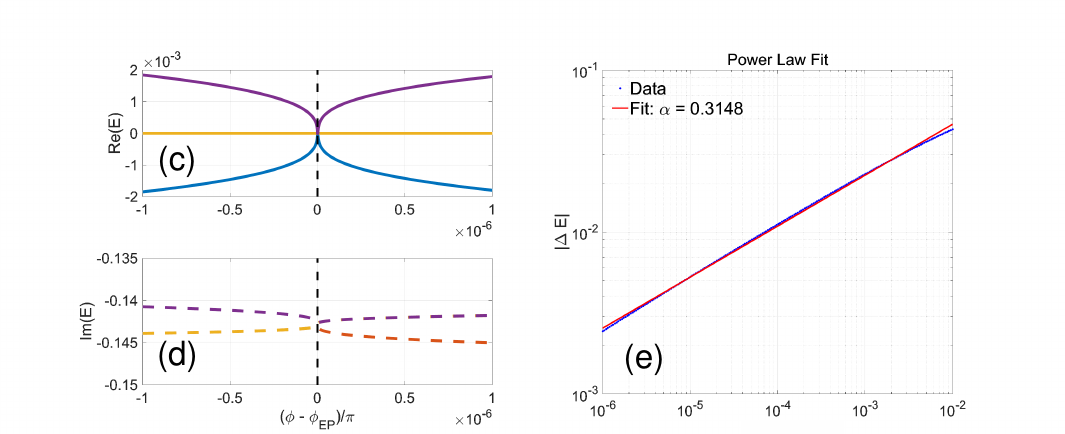}
		\end{minipage}
		\caption{(Color online) (a) Complex-energy spectrum for $\Gamma=0.25$, $\gamma_\uparrow=0.75$, $\gamma_\downarrow=0.25$, $\phi_\mathbf{B}=0.24075202\pi$, $t_d=0$, and $\Delta\to\infty$. Solid and dashed curves denote real and imaginary parts, respectively. (b) Corresponding five-level finite-gap surrogate spectrum at $\Delta=1$. (c),(d) Real and imaginary parts of the eigenvalue splitting near $\phi_{\mathrm{EP}}$. (e) Power-law fit of $\Delta E$ versus $|\delta\phi-\phi_{\mathrm{EP}}|$.}
	\end{figure*}
	
	At the refined parameter values $\phi_{\mathbf B}=0.24075202\pi$ and $\delta\phi=0.51884305\pi\equiv\phi_{\mathrm{EP}}$, two second-order EPs merge at $E_{\mathrm{EP}}=-0.142849,\mathrm{i}$ in the infinite-gap model.
	The numerical splitting follows the scaling behavior $\Delta E\propto|\delta\phi-\phi_{\mathrm{EP}}|^{\alpha}$ with $\alpha=0.3148\simeq1/3$ [Fig.~4(e)]. 
	We additionally analyzed the null spaces of $A=\mathcal H-E_{\mathrm{EP}}$: with a relative singular-value tolerance of $10^{-10}$, $\dim\ker A=1$, $\dim\ker A^2=2$, and $\dim\ker A^3=3$. 
	This length-three Jordan chain directly identifies a third-order EP.
	At the same flux and phase values, the $\Delta=1$ five-level surrogate model does not preserve the triple coalescence; 
	this parameter-specific result should not be interpreted as a general mechanism for eliminating third-order EPs.
	
	\subsection{Josephson current and multi-particle representation}
	The equilibrium Josephson current is obtained from the phase derivative of the free energy:
	\begin{equation}
		I_\mathrm{J}=2\frac{\partial F}{\partial\,\delta\phi}.
		\label{Free_Energy_Current}
	\end{equation}
	We concentrate on the spin-selective regime where second-order EPs persist in the finite-gap superconducting system.
	
	\begin{figure*}[htbp]
		\begin{minipage}{0.76\linewidth}
			\centering
			\makebox[0.48\linewidth][c]{\(\phi_B=0.5\pi\)}
			\makebox[0.48\linewidth][c]{\(\phi_B=0.240753\pi\)}\\[-0.4ex]
			\includegraphics[width=0.48\linewidth]{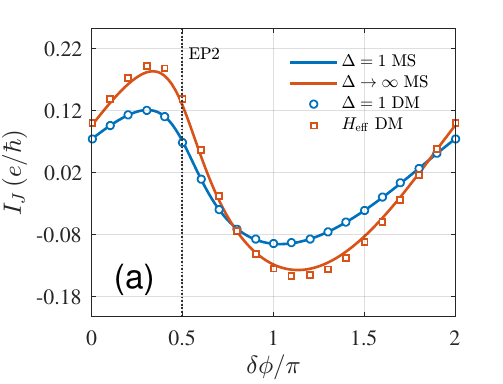}
			\includegraphics[width=0.48\linewidth]{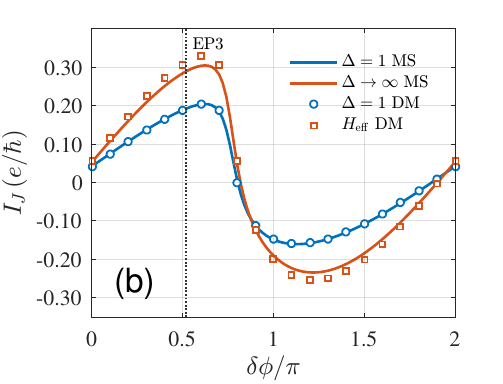}
		\end{minipage}
		\caption{
			(Color online) Josephson current at the two magnetic fluxes indicated above the panels.
			The currents are calculated using the Matsubara free-energy derivative
			\(I_J=2\partial F/\partial\delta\phi\), and the density-matrix expectation value
			\(I_J=2\langle \partial \tilde{H}/\partial\delta\phi\rangle\).
			In the legend, MS and DM denote the Matsubara-summation and density-matrix calculations, respectively.
			For the finite-gap surrogate Hamiltonian with \(\Delta=1\), the two methods give consistent results.
			The large-gap result with \(\Delta=100\) approaches the infinite-gap low-energy effective model.
			The vertical dotted lines mark the EP2 and EP3 phase positions identified in Figs.~3(a) and 4(a), respectively; the EP3 label denotes the infinite-gap coalescence, which is lifted at \(\Delta=1\).
			Common parameters are \(\Gamma=0.25\), \(t_d=0\), and \(\Delta=1\) for the finite-gap results.
		}
		\label{fig_current}
	\end{figure*}
	
	Compared with the large-gap limit, the finite-gap surrogate model reduces the current amplitude and modifies the current-phase relationship.  
	The magnetic flux shifts this relationship by altering the interference between the two dot paths. 
	Notably, the current remains continuous as the phase passes through the exceptional points, consistent with the distinction between spectral nonanalyticity and thermodynamic response in non-Hermitian Josephson junctions~\cite{Pino2025,Shen2024}.
	We compare the free-energy result with the density-matrix method introduced for non-Hermitian equilibrium transport in Ref.~\cite{Shen2024}. 
	In the surrogate Hamiltonian framework, this current is expressed as:
	\begin{figure*}[t]
		\centering
		\begin{minipage}{0.64\linewidth}
			\centering
			\makebox[0.48\linewidth]{Odd parity}
			\makebox[0.48\linewidth]{Even parity}\\[-0.2ex]
			\includegraphics[width=0.48\linewidth]{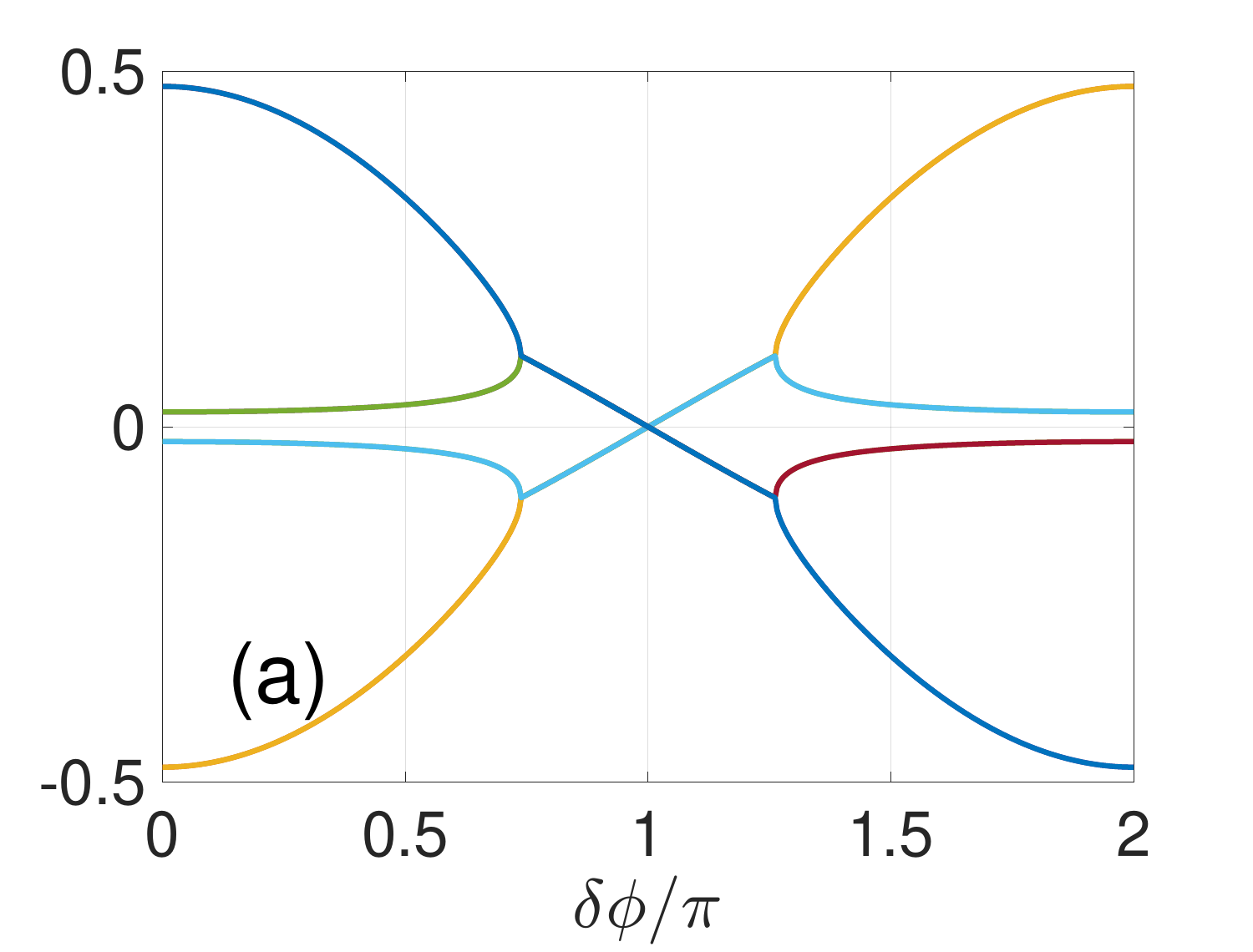}
			\includegraphics[width=0.48\linewidth]{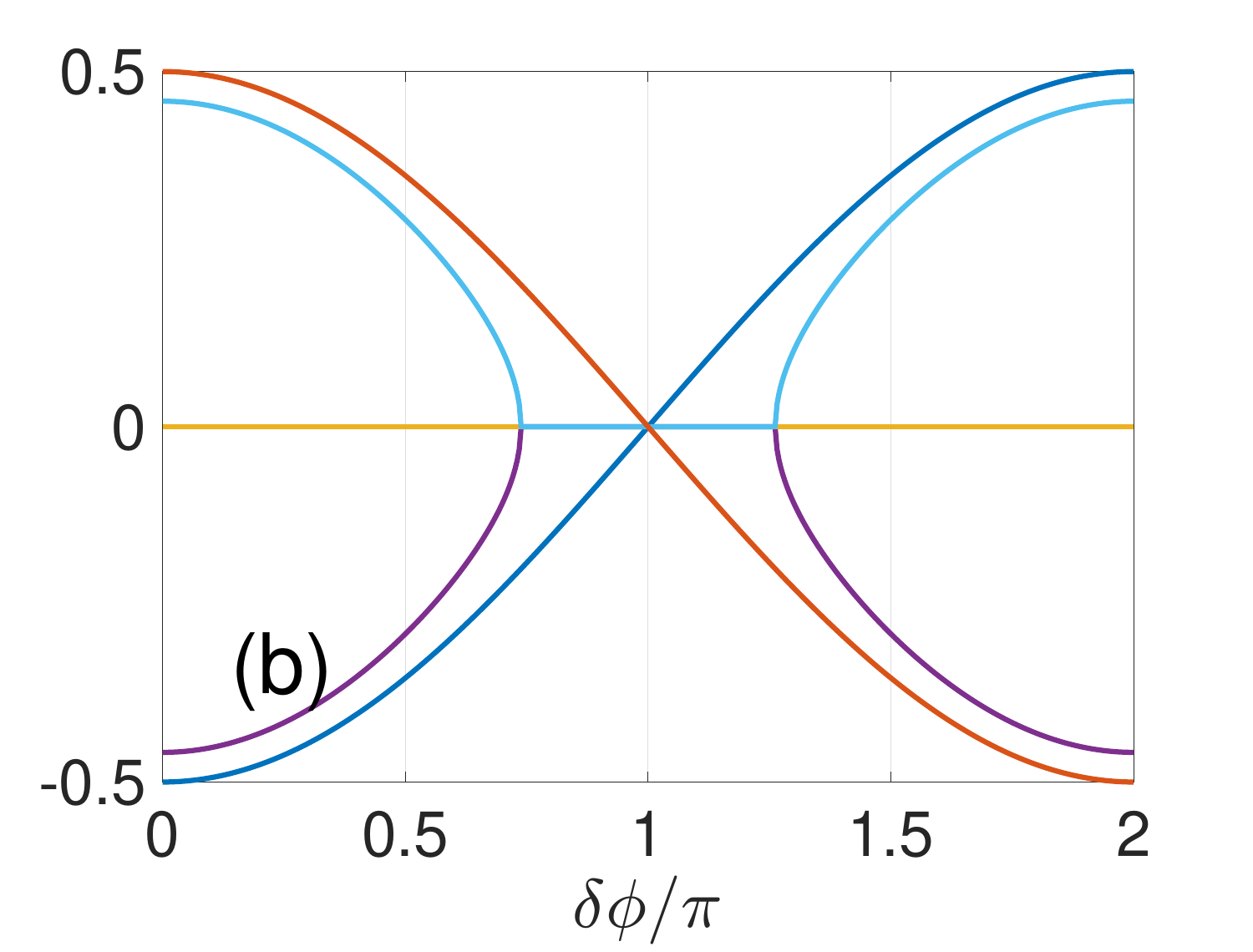}
			\includegraphics[width=0.48\linewidth]{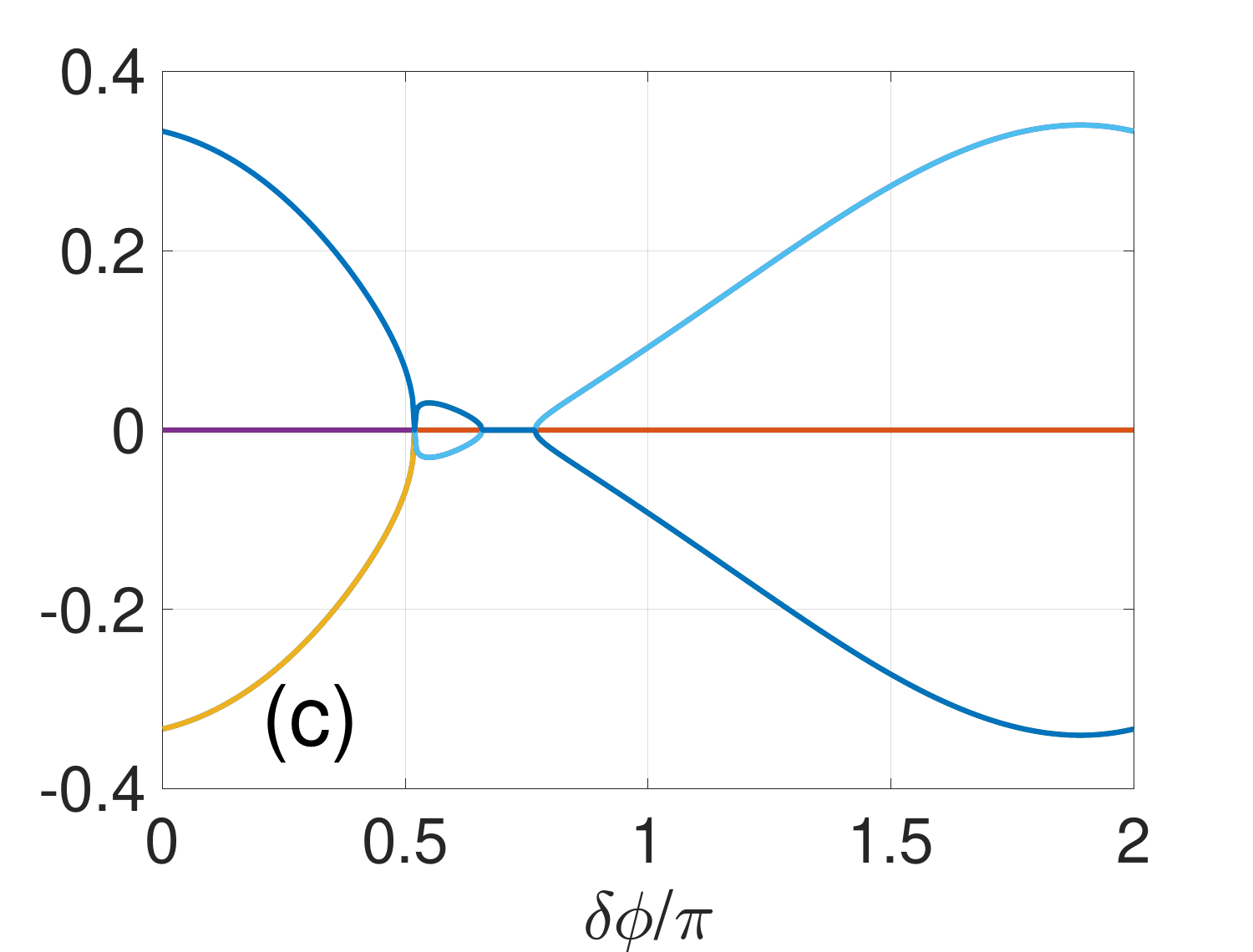}
			\includegraphics[width=0.48\linewidth]{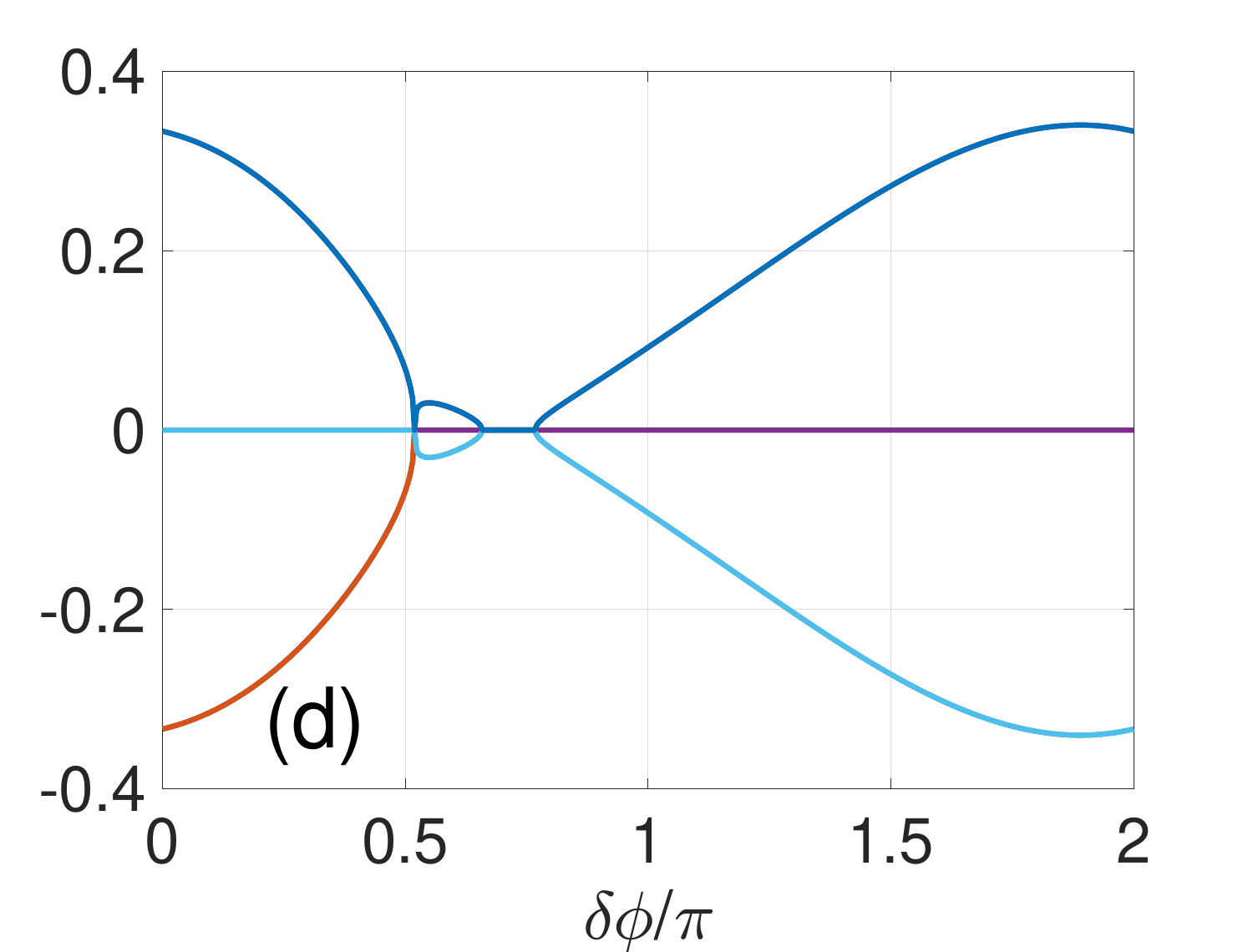}
		\end{minipage}
		\caption{(Color online) Real part of energy spectrum from parallel double quantum dot system based on the $H_{\mathrm{eff}}$ Hamiltonian in multi-particle representation. The left and right columns show the odd- and even-parity sectors, respectively. The top row corresponds to the EP realization in Fig.~2(a): $\Gamma = 0.25$, $\gamma_\uparrow = \gamma_\downarrow = 0.2$, $\phi_\mathbf{B} = 0$. The bottom row corresponds to the EP realization in Fig.~4(a): $\Gamma = 0.25$, $\gamma_\uparrow = 0.75, \gamma_\downarrow = 0.25$, $\phi_\mathbf{B} = 0.240753\,\pi$. Other parameters are set as $t_d=0$, $\Delta\to\infty$.}
		\label{fig_parity_spectrum}
	\end{figure*}

	\begin{align}
	I_\mathrm{J}&=2\left\langle
	\frac{\partial \tilde H}{\partial\delta\phi}\right\rangle \nonumber\\
	&=2\Delta\,\mathrm{Im}\!\left[
	\sum_{l=-2}^{2}\langle c^\mathrm{L}_{l\uparrow}c^\mathrm{L}_{l\downarrow}\rangle
	\mathrm{e}^{-\mathrm{i}\delta\phi/2}\right] \nonumber\\
	&\quad-2\Delta\,\mathrm{Im}\!\left[
	\sum_{l=-2}^{2}\langle c^\mathrm{R}_{l\uparrow}c^\mathrm{R}_{l\downarrow}\rangle
	\mathrm{e}^{\mathrm{i}\delta\phi/2}\right].
	\end{align}
	where the positive-current direction is defined by $I_J=2\partial F/\partial\delta\phi$, and $\langle c^\eta_{l\uparrow}c^\eta_{l\downarrow}\rangle$ is evaluated using the biorthogonal single-particle density matrix of the retarded surrogate Hamiltonian according to Ref.~\cite{Shen2024}. 
	The dotted curves in Fig.~5 demonstrate agreement with the Matsubara free-energy result at $\Delta=1$. 
	At $\Delta=100$, both calculations converge toward the infinite-gap effective-model result. 
	This agreement serves as a numerical consistency check of the finite-gap surrogate construction; however, it does not, by itself, establish a general equivalence between the two formalisms.

	To identify the origin of the contrasting finite-gap behavior, we also express $H_{\mathrm{eff}}$ in the many-body Fock space. The 16-dimensional Hilbert space separates into two 8-dimensional blocks of even and odd fermion parity. We diagonalize both blocks for the parameter sets of Figs.~2(a) and 4(a), as shown in Fig.~\ref{fig_parity_spectrum}.
	
	For the EP realization depicted in Fig.~2(a) [Figs.~\ref{fig_parity_spectrum}(a) and \ref{fig_parity_spectrum}(b)], the real parts of the even- and odd-parity spectra exhibit distinct structural characteristics. 
	For the realization shown in Fig.~4(a) [Figs.~\ref{fig_parity_spectrum}(c) and \ref{fig_parity_spectrum}(d)], they instead demonstrate a matching pattern. 
	This contrast correlates with, and provides a potential microscopic interpretation of, the different finite-gap behavior. 
	However, it does not, by itself, establish symmetry protection: such an analysis would require matching the complete complex spectra and identifying an explicit symmetry or similarity transformation between the parity blocks. 
	The parity-resolved analysis therefore complements the Green's-function separation criterion of Ref.~\cite{Capecelatro2025} without substituting for it.
	
	\section{Conclusion}
	We have investigated EPs in a parallel double-quantum-dot Josephson junction coupled to dissipative normal or ferromagnetic reservoirs. 
	Within the infinite-gap approximation, spin-independent loss generates second-order EPs. 
	The five-level finite-gap surrogate model replaces these eigenvalue coalescences with avoided crossings for the parameters examined. 
	With spin-dependent loss and orbital flux, a subset of second-order EPs persists, and direct comparison with the continuous finite-gap self-energy confirms the relevant EP shown in Fig.~3 with only a minor phase displacement. 
	Flux tuning further merges two singularities into a third-order EP exhibiting cubic-root splitting in the infinite-gap model; however, the same fine-tuned parameters do not yield a triple coalescence in the $\Delta=1$ surrogate.
	
	The many-body analysis indicates a correlation: surrogate coalescences that disappear and those that persist exhibit distinct relationships between their parity-resolved real-energy patterns. Establishing this correlation as a protection mechanism would necessitate a symmetry analysis of the complete complex spectra. 
	The Josephson current remains continuous at the second-order EPs that survive in the surrogate model. 
	The present results therefore identify spin-selective dissipation and interferometric flux as effective control parameters for non-Hermitian degeneracies, while distinguishing directly validated continuous-model conclusions from parameter-specific surrogate predictions.
\clearpage
\bibliography{ref}

@article{Okuma2023,
  title={Non-Hermitian Topological Phenomena: A Review},
  author={Okuma, Nobuyuki and Sato, Masatoshi},
  journal={Annual Review of Condensed Matter Physics},
  volume={14},
  pages={83--107},
  year={2023},
  publisher={Annual Reviews}
}

@article{Shen2024,
  title = {Non-Hermitian Fermi-Dirac Distribution in Persistent Current Transport},
  author = {Shen, Pei-Xin and Lu, Zhide and Lado, Jose L. and Trif, Mircea},
  journal = {Phys. Rev. Lett.},
  volume = {133},
  issue = {8},
  pages = {086301},
  numpages = {8},
  year = {2024},
  month = {Aug},
  publisher = {American Physical Society},
  doi = {10.1103/PhysRevLett.133.086301},
  url = {https://link.aps.org/doi/10.1103/PhysRevLett.133.086301}
}

@article{Pino2025,
  title={Thermodynamics of non-Hermitian Josephson junctions with exceptional points},
  author={Pino, D. Michel and Meir, Yigal and Aguado, Ram{\'o}n},
  journal={Physical Review B},
  volume={111},
  number={14},
  pages={L140503},
  year={2025},
  publisher={American Physical Society}
}

@article{Solow2025,
  title = {Signatures of exceptional points in multiterminal superconductor--normal metal junctions},
  author = {Solow, Oliver and Flensberg, Karsten},
  journal = {Phys. Rev. B},
  volume = {112},
  issue = {16},
  pages = {L161402},
  numpages = {5},
  year = {2025},
  month = {Oct},
  publisher = {American Physical Society},
  doi = {10.1103/dmfm-71l6},
  url = {https://link.aps.org/doi/10.1103/dmfm-71l6}
}

@article{Capecelatro2025,
  title = {Andreev non-Hermitian Hamiltonian for open Josephson junctions from Green's functions},
  author = {Capecelatro, R. and Marciani, M. and Campagnano, G. and Lucignano, P.},
  journal = {Phys. Rev. B},
  volume = {111},
  issue = {6},
  pages = {064517},
  numpages = {23},
  year = {2025},
  month = {Feb},
  publisher = {American Physical Society},
  doi = {10.1103/PhysRevB.111.064517},
  url = {https://link.aps.org/doi/10.1103/PhysRevB.111.064517}
}

@article{Kawabata2019,
  author  = {Kawabata, Kohei and Bessho, Takumi and Sato, Masatoshi},
  title   = {Classification of Exceptional Points and Non-Hermitian Topological Semimetals},
  journal = {Physical Review Letters},
  volume  = {123},
  number  = {6},
  pages   = {066405},
  year    = {2019},
  doi     = {10.1103/PhysRevLett.123.066405}
}

@article{Bessho2020,
  author  = {Bessho, Takumi and Kawabata, Kohei and Sato, Masatoshi},
  title   = {Topological Classification of Non-Hermitian Gapless Phases: Exceptional Points and Bulk Fermi Arcs},
  journal = {Journal of the Physical Society of Japan Conference Proceedings},
  volume  = {30},
  pages   = {011098},
  year    = {2020},
  doi     = {10.7566/JPSCP.30.011098}
}

@article{Ding2022,
  author  = {Ding, Kun and Fang, Chen and Ma, Guancong},
  title   = {Non-Hermitian Topology and Exceptional-Point Geometries},
  journal = {Nature Reviews Physics},
  volume  = {4},
  number  = {12},
  pages   = {745--760},
  year    = {2022},
  doi     = {10.1038/s42254-022-00516-5}
}

@article{KoziiFu2024,
  author  = {Kozii, Vladyslav and Fu, Liang},
  title   = {Non-Hermitian Topological Theory of Finite-Lifetime Quasiparticles},
  journal = {Physical Review B},
  volume  = {109},
  number  = {23},
  pages   = {235139},
  year    = {2024},
  doi     = {10.1103/PhysRevB.109.235139}
}

@article{Cayao2024PhaseBiased,
  author  = {Cayao, Jorge and Sato, Masatoshi},
  title   = {Non-Hermitian Phase-Biased Josephson Junctions},
  journal = {Physical Review B},
  volume  = {110},
  number  = {20},
  pages   = {L201403},
  year    = {2024},
  doi     = {10.1103/PhysRevB.110.L201403}
}

@article{Cayao2024MultiTerminal,
  author  = {Cayao, Jorge and Sato, Masatoshi},
  title   = {Non-Hermitian Multiterminal Phase-Biased Josephson Junctions},
  journal = {Physical Review B},
  volume  = {110},
  number  = {23},
  pages   = {235426},
  year    = {2024},
  doi     = {10.1103/PhysRevB.110.235426}
}

@article{Ohnmacht2025,
  author  = {Ohnmacht, David Christian and Wilhelm, Valentin and Weisbrich, Hannes and Belzig, Wolfgang},
  title   = {Non-Hermitian Topology in Multiterminal Superconducting Junctions},
  journal = {Physical Review Letters},
  volume  = {134},
  number  = {15},
  pages   = {156601},
  year    = {2025},
  doi     = {10.1103/PhysRevLett.134.156601}
}

@article{MartinRodero2011,
  author  = {Mart{\'i}n-Rodero, A. and Levy Yeyati, A.},
  title   = {Josephson and Andreev Transport through Quantum Dots},
  journal = {Advances in Physics},
  volume  = {60},
  number  = {6},
  pages   = {899--958},
  year    = {2011},
  doi     = {10.1080/00018732.2011.624266}
}

@article{Probst2016,
  author  = {Probst, B. and Dom{\'i}nguez, F. and Schroer, A. and Levy Yeyati, A. and Recher, P.},
  title   = {Signatures of Nonlocal Cooper-Pair Transport and of a Singlet-Triplet Transition in the Critical Current of a Double-Quantum-Dot Josephson Junction},
  journal = {Physical Review B},
  volume  = {94},
  pages   = {155445},
  year    = {2016},
  doi     = {10.1103/PhysRevB.94.155445}
}

@article{Tomaszewski2018,
  author  = {Tomaszewski, D. and Busz, P. and L{\'o}pez, R. and Lee, M. and Martinek, J.},
  title   = {Aharonov--Bohm and Aharonov--Casher Effects in a Double Quantum Dot Josephson Junction},
  journal = {Physical Review B},
  volume  = {98},
  pages   = {174504},
  year    = {2018},
  doi     = {10.1103/PhysRevB.98.174504}
}

@article{Bergholtz2021,author={Bergholtz, E. J. and Budich, J. C. and Kunst, F. K.},title={Exceptional topology of non-Hermitian systems},journal={Reviews of Modern Physics},volume={93},pages={015005},year={2021},doi={10.1103/RevModPhys.93.015005}}

@article{Ashida2021,author={Ashida, Y. and Gong, Z. and Ueda, M.},title={Non-Hermitian physics},journal={Advances in Physics},volume={69},pages={249--435},year={2021},doi={10.1080/00018732.2021.1876991}}

@article{ElGanainy2018,author={El-Ganainy, R. and Makris, K. G. and Khajavikhan, M. and Musslimani, Z. H. and Rotter, S. and Christodoulides, D. N.},title={Non-Hermitian physics and PT symmetry},journal={Nature Physics},volume={14},pages={11--19},year={2018},doi={10.1038/nphys4323}}

@article{Feng2017,author={Feng, L. and El-Ganainy, R. and Ge, L.},title={Non-Hermitian photonics based on parity-time symmetry},journal={Nature Photonics},volume={11},pages={752--762},year={2017},doi={10.1038/s41566-017-0031-1}}

@article{Dembowski2001,author={Dembowski, C. and Gr{"a}f, H.-D. and Harney, H. L. and Heine, A. and Heiss, W. D. and Rehfeld, H. and Richter, A.},title={Experimental observation of the topological structure of exceptional points},journal={Physical Review Letters},volume={86},pages={787--790},year={2001},doi={10.1103/PhysRevLett.86.787}}

@article{Xu2016,author={Xu, H. and Mason, D. and Jiang, L. and Harris, J. G. E.},title={Topological energy transfer in an optomechanical system with exceptional points},journal={Nature},volume={537},pages={80--83},year={2016},doi={10.1038/nature18604}}

@article{Doppler2016,author={Doppler, J. and Mailybaev, A. A. and B{"o}hm, J. and Kuhl, U. and Girschik, A. and Libisch, F. and Milburn, T. J. and Rabl, P. and Moiseyev, N. and Rotter, S.},title={Dynamically encircling an exceptional point for asymmetric mode switching},journal={Nature},volume={537},pages={76--79},year={2016},doi={10.1038/nature18605}}

@article{Heiss2012,author={Heiss, W. D.},title={The physics of exceptional points},journal={Journal of Physics A},volume={45},pages={444016},year={2012},doi={10.1088/1751-8113/45/44/444016}}

@article{Rotter2009,author={Rotter, I.},title={A non-Hermitian Hamilton operator and the physics of open quantum systems},journal={Journal of Physics A},volume={42},pages={153001},year={2009},doi={10.1088/1751-8113/42/15/153001}}

@article{Miri2019,author={Miri, M.-A. and Al{\`u}, A.},title={Exceptional points in optics and photonics},journal={Science},volume={363},pages={eaar7709},year={2019},doi={10.1126/science.aar7709}}

@article{Buzdin2005,author={Buzdin, A. I.},title={Proximity effects in superconductor-ferromagnet heterostructures},journal={Reviews of Modern Physics},volume={77},pages={935--976},year={2005},doi={10.1103/RevModPhys.77.935}}

@article{Bergeret2005,author={Bergeret, F. S. and Volkov, A. F. and Efetov, K. B.},title={Odd triplet superconductivity and related phenomena in superconductor-ferromagnet structures},journal={Reviews of Modern Physics},volume={77},pages={1321--1373},year={2005},doi={10.1103/RevModPhys.77.1321}}

@article{Linder2015,author={Linder, J. and Robinson, J. W. A.},title={Superconducting spintronics},journal={Nature Physics},volume={11},pages={307--315},year={2015},doi={10.1038/nphys3242}}

@article{Beenakker1991,author={Beenakker, C. W. J.},title={Universal limit of critical-current fluctuations in mesoscopic Josephson junctions},journal={Physical Review Letters},volume={67},pages={3836--3839},year={1991},doi={10.1103/PhysRevLett.67.3836}}

@article{Hays2018,author={Hays, M. and de Lange, G. and Serniak, K. and van Woerkom, D. J. and Bouman, D. and Krogstrup, P. and Nyg{\aa}rd, J. and Geresdi, A. and Devoret, M. H.},title={Direct microwave measurement of Andreev-bound-state dynamics in a semiconductor-nanowire Josephson junction},journal={Physical Review Letters},volume={121},pages={047001},year={2018},doi={10.1103/PhysRevLett.121.047001}}

@article{Pillet2010,author={Pillet, J.-D. and Quay, C. H. L. and Morfin, P. and Bena, C. and Levy Yeyati, A. and Joyez, P.},title={Andreev bound states in supercurrent-carrying carbon nanotubes revealed},journal={Nature Physics},volume={6},pages={965--969},year={2010},doi={10.1038/nphys1811}}

@article{vanDam2006,author={van Dam, J. A. and Nazarov, Y. V. and Bakkers, E. P. A. M. and De Franceschi, S. and Kouwenhoven, L. P.},title={Supercurrent reversal in quantum dots},journal={Nature},volume={442},pages={667--670},year={2006},doi={10.1038/nature05018}}

@article{Eichler2007,author={Eichler, A. and Weiss, M. and Schafhaeutle, C. and Moser, J. and Schonenberger, C.},title={Even-odd effect in Andreev transport through a carbon nanotube quantum dot},journal={Physical Review Letters},volume={99},pages={126602},year={2007},doi={10.1103/PhysRevLett.99.126602}}

@article{Karrasch2008,author={Karrasch, C. and Oguri, A. and Meden, V.},title={Josephson current through a single Anderson impurity coupled to BCS leads},journal={Physical Review B},volume={77},pages={024517},year={2008},doi={10.1103/PhysRevB.77.024517}}

@article{Zitko2010,author={Zitko, R. and Lee, M. and L{\'o}pez, R. and Aguado, R. and Choi, M.-S.},title={Josephson current in strongly correlated double quantum dots},journal={Physical Review Letters},volume={105},pages={116803},year={2010},doi={10.1103/PhysRevLett.105.116803}}

@article{Choi2000,author={Choi, M.-S. and Bruder, C. and Loss, D.},title={Spin-dependent Josephson current through double quantum dots and measurement of entangled electron states},journal={Physical Review B},volume={62},pages={13569--13572},year={2000},doi={10.1103/PhysRevB.62.13569}}

@article{Wang2025Flux,
  author = {Wang, Yiyan and Li, Cong and Dong, Bing},
  title = {Magnetic flux controlled current phase relationship in double quantum dot Josephson junction},
  journal = {New Journal of Physics},
  volume = {28},
  pages = {043506},
  year = {2026},
  doi = {10.1088/1367-2630/ae59ab}
}

@article{Li2025QPT,
  author = {Li, Cong and Wang, Yiyan and Dong, Bing},
  title = {Quantum phase transition in a double quantum dot Josephson junction driven by electron-electron interactions},
  journal = {Physical Review B},
  volume = {112},
  pages = {165430},
  year = {2025},
  doi = {10.1103/b4tp-7x57}
}
\end{document}